\documentclass[sigconf]{acmart}
\AtBeginDocument{%
  \providecommand\BibTeX{{%
    \normalfont B\kern-0.5em{\scshape i\kern-0.25em b}\kern-0.8em\TeX}}}

\setcopyright{acmcopyright}
\copyrightyear{2018}
\acmYear{2018}
\acmDOI{XXXXXXX.XXXXXXX}

\acmPrice{15.00}
\acmISBN{978-1-4503-XXXX-X/18/06}

\usepackage{afterpage}
\usepackage{booktabs}
\usepackage{multirow}
\usepackage{algorithm}
\usepackage{algorithmic}
\usepackage{enumitem}
\setlist[itemize]{leftmargin=*}

\newcommand{\etal}{\textit{et al}. }

\newcommand{\eg}{\textit{e.g.}}

\newcommand{\ie}{\emph{i.e.}}

\renewcommand\footnotetextcopyrightpermission[1]{}

\begin{document}
% \fancyhead{}
% \title{Large Language Model for Preference Reasoning}
% \title{Edge Integrated Re-ranking with Real-time Heterogeneous User Behavior Modeling}
% \title{Deepen User Insights: Leveraging Large Language Models for Augmenting Representations in Open-world Recommendation}
\title{Towards Open-World Recommendation with Knowledge Augmentation from Large Language Models}
% \title{Understand You Better: Leveraging Open-world Knowledge for Representation Augmented Recommendation with Large Language Models}
% \title{Towards Open-world Recommender Systems with Large Language Models}
% Bridging Large Language Models and Recommender Systems: A xxx Approach
% Deepen User Insights: 
% REALLM Representation augmented recommendation with large language models.
% LLM4xxx

\author{Yunjia Xi}
\authornote{Both authors contributed equally to this research.}
\email{xiyunjia@sjtu.edu.cn}
\affiliation{%
  \institution{Shanghai Jiao Tong University}
  \city{Shanghai}
  \country{China}
}

\author{Weiwen Liu}
\authornotemark[1]
\email{liuweiwen8@huawei.com}
\affiliation{%
  \institution{Huawei Noah's Ark Lab}
  \city{Shenzhen}
  \country{China}
}

\author{Jianghao Lin}
\email{chiangel@sjtu.edu.cn}
\affiliation{%
  \institution{Shanghai Jiao Tong University}
  \city{Shanghai}
  \country{China}
}

\author{Xiaoling Cai}
\email{caixiaoling2@huawei.com}
\affiliation{%
  \institution{Consumer Business Group, Huawei}
  \city{Shenzhen}
  \country{China}
}

\author{Hong Zhu}
\email{zhuhong8@huawei.com}
\affiliation{%
  \institution{Consumer Business Group, Huawei}
  \city{Shenzhen}
  \country{China}
}

\author{Jieming Zhu}
\email{jamie.zhu@huawei.com}
\affiliation{%
  \institution{Huawei Noah's Ark Lab}
  \city{Shenzhen}
  \country{China}
}

\author{Bo Chen}
\email{chenbo116@huawei.com}
\affiliation{%
  \institution{Huawei Noah's Ark Lab}
  \city{Shenzhen}
  \country{China}
}

\author{Ruiming Tang}
\email{tangruiming@huawei.com}
\affiliation{%
  \institution{Huawei Noah's Ark Lab}
  \city{Shenzhen}
  \country{China}
}

\author{Weinan Zhang}
\email{wnzhang@sjtu.edu.cn}
\affiliation{%
  \institution{Shanghai Jiao Tong University}
  \city{Shanghai}
  \country{China}
}

\author{Rui Zhang}
\email{rayteam@yeah.net}
\affiliation{%
  \institution{ruizhang.info}
  \city{Shenzhen}
  \country{China}
}

\author{Yong Yu}
\email{yyu@sjtu.edu.cn}
\affiliation{%
  \institution{Shanghai Jiao Tong University}
  \city{Shanghai}
  \country{China}
}

\renewcommand{\shortauthors}{Yunjia Xi et al.}

\begin{abstract}
Recommender systems play a vital role in various online services. However, the insulated nature of training and deploying separately within a specific domain limits their access to open-world knowledge. Recently, the emergence of large language models (LLMs) has shown promise in bridging this gap by encoding extensive world knowledge and demonstrating reasoning capability. Nevertheless, previous attempts to directly use LLMs as recommenders have not achieved satisfactory results. In this work, we propose an Open-World \underline{K}nowledge \underline{A}ugmented \underline{R}ecommendation Framework with Large Language Models, dubbed \textit{KAR}, to acquire two types of external knowledge from LLMs --- the \textit{reasoning knowledge} on user preferences and the \textit{factual knowledge} on items. We introduce \textit{factorization prompting} to elicit accurate reasoning on user preferences. The generated reasoning and factual knowledge are effectively transformed and condensed into augmented vectors by a \textit{hybrid-expert adaptor} in order to be compatible with the recommendation task. The obtained vectors can then be directly used to enhance the performance of any recommendation model. We also ensure efficient inference by preprocessing and prestoring the knowledge from the LLM. Extensive experiments show that KAR significantly outperforms the state-of-the-art baselines and is compatible with a wide range of recommendation algorithms. We deploy KAR to Huawei's news and music recommendation platforms and gain a 7\% and 1.7\% improvement in the online A/B test, respectively.

\end{abstract}

% \cb{, which is ...}



% \keywords{Recommender System, Edge Computing, Integrated Re-ranking}

\maketitle
\section{Introduction}
% recommender system are closed systems, lack world knowledge
% capability of large lanugage models: reasoning and world knowledge; black box and white box llm; Adopting llm for rec as in existing methods [xxx] is infeasible: a) lack rec domain knowledge, cannot fast inference according to ever-changing user behavior data; b) require extreme large computational resource to train the model.

% fully exploit reasoning and world knowledge of llm as well as domain knowledge, while keep efficient inference.
% challenge: the utilization of llm to recsys: a) text/super large dense vector b) noise information c) in syntactic and recommendation space are not aligned (unisrec).

\begin{figure}
    \centering
    \includegraphics[width=0.48\textwidth]{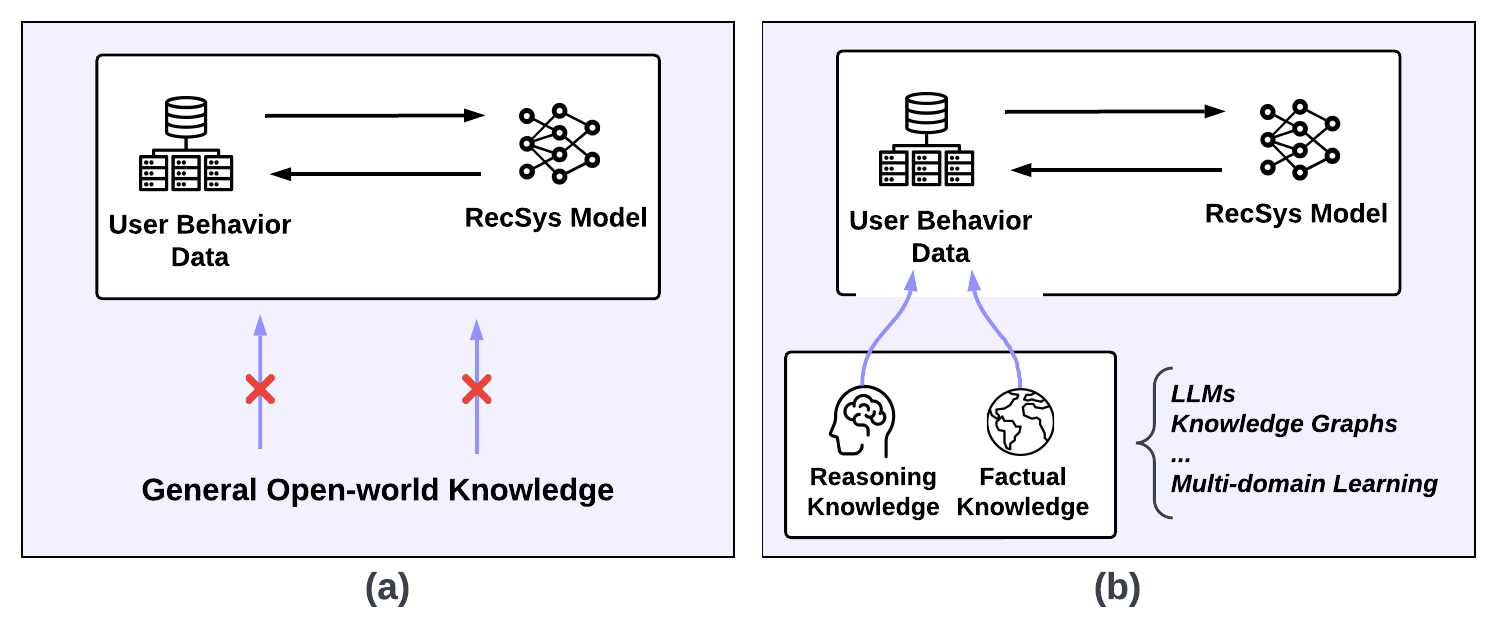}
    % \caption{Comparison between closed recommender systems and open-world recommender systems. (a) Closed recommender systems. (b) Open-world recommender systems.
    \vspace{-20pt}
    \caption{Comparison between (a) closed recommender systems and (b) open-world recommender systems.}
    \vspace{-15pt}
    % \Jianghao{Comparison between (a) closed recommender systems and (b) open-world recommender systems.}}
    \label{fig:intro}
\end{figure}

Recommender systems (RSs) are ubiquitous in today's online services, shaping and enhancing user experiences in various domains such as movie discovery \cite{koren2009matrix}, online shopping \cite{he2016ups}, and music streaming~\cite{van2013deep}. However, a common characteristic of existing recommender systems is their \textit{insulated nature} --- the models are trained and deployed within closed systems.

As depicted in Figure~\ref{fig:intro}(a), the data utilized in a classical recommender system is confined to one or a few specific application domains \cite{koren2009matrix, DIN}, isolated from the knowledge of the external world, thereby restricting the information that could be learned for a recommendation model. In fact, knowledge beyond the given domains can significantly enhance the predictive accuracy and the generalization ability of recommender systems \cite{friedman2023leveraging,lin2023sparks}. Hence, in this work, we posit that instead of solely learning from narrowly defined data in the \textbf{closed systems}, recommender systems should be the \textbf{open-world systems} that can proactively acquire knowledge from the external world, as shown in Figure~\ref{fig:intro}(b).

In particular, two types of information from the external world are especially useful for recommendation, which we refer to as \textit{open-world knowledge} for recommendation --- the \textit{reasoning knowledge} on in-depth user preferences which is inferred from user behaviors and profiles, and the \textit{factual knowledge} on items that can be directly obtained from the web. On the one hand, the reasoning knowledge inferred from the user behavior history enables a more comprehensive understanding of the users, and is critical for better recommendation performance. Deducing the underlying preferences and motives that drive user behaviors can help us gain deeper insights and clues about the users. A person's personality, occupation, intentions, preferences, and tastes could be reflected in their behavior history.  This preference reasoning can even integrate seasonal factors (\eg, holiday-themed movie preferences during Christmas) or external events (\eg, an increased interest in health products during a pandemic) and provide human-like recommendations with clear evidence, which goes beyond identifying basic behavior patterns as in classical recommenders. On the other hand, the factual knowledge on items provides valuable common sense information about the candidate items and thereby improves the recommendation quality. Take movie recommendation as an example, the external world contains additional movie features such as \textit{plots, related reports, awards, critic reviews} that have not been included in the recommendation dataset, which expands the original data and is beneficial to the recommendation task. 

% \cb{just for user preference reasoning? item relations reasoning?}

Several existing studies attempt to complement the closed recommender systems with additional information by knowledge graphs~\cite{guo2020survey,wang2019knowledge} or multi-domain learning~\cite{star,guo2023disentangled}. However, constructing comprehensive and accurate knowledge graphs or multi-domain datasets requires considerable extra human effort, and the accessible knowledge remains limited. Moreover, they only focus on extracting the factual knowledge from the external world, and overlook the reasoning knowledge on user preferences~\cite{guo2020survey}.  For instance, knowledge graphs used in RS usually focus on complement knowledge for items and rarely consider users~\cite{guo2020survey,wang2019knowledge,wang2019multi} since the user-side knowledge is highly dynamic and challenging to capture within a fixed knowledge graph.

Recent rapid developments in large language models (LLMs) have revolutionized the learning paradigm of various research fields and show great potential in bridging the gap between classical recommenders and open-world knowledge \cite{zhao2023survey}. With the immense scale of the model and the corpus size, these large pretrained language models like GPT-4 \cite{gpt4}, LLaMA \cite{llama} have shown remarkable capabilities, such as problem solving, logical reasoning, creative writing~\cite{bubeck2023sparks}. Learning from an extensive corpus of internet texts, LLMs have encoded a vast array of world knowledge --- from basic factual information to complex societal norms and logical structures \cite{bubeck2023sparks}. As a result, LLMs can perform basic logical reasoning that aligns with known facts and relationships \cite{wei2022chain,zhou2023leasttomost}. 

Recently, a few studies have attempted to apply LLMs as recommenders by converting recommendation tasks and user profiles into prompts \cite{chatrec,liu2023chatgpt,dai2023uncovering}. Though some preliminary findings have been obtained, the results of using LLMs as recommenders are far from optimal for real-world recommendation scenarios due to the following shortcomings. 1) \textbf{Predictive accuracy.} The accuracy of LLMs is generally outperformed by classical recommenders in most cases, since LLMs have not been trained on specific recommendation data \cite{liu2023chatgpt}. The lack of recommendation domain knowledge and collaborative signals prevents LLMs from adapting to individual user preferences. 
2)~\textbf{Inference latency.} Due to the excessive number of model parameters, it is impractical to directly use LLMs as recommender systems in industrial settings. With billions of users and thousands of user behaviors, LLMs fail to meet the low latency requirement in recommender systems (usually within 100 milliseconds). The large model size also hinders the possibility of employing real-time user feedback to update and refine the model as in classical recommenders.
3)~\textbf{Compositional gap.} LLMs often suffer from the issue of compositional gap, where LLMs have difficulty in generating correct answers to the compositional problem like recommending items to users, whereas they can correctly answer all its sub-problems \cite{press2022measuring}. Requiring direct recommendation results from LLMs is currently beyond their capability and cannot fully exploit the open-world knowledge encoded in LLMs \cite{kang2023llms,dai2023uncovering}.

 % \Jianghao{Model complexity is also the reason why not tune LLM. I think \textbf{Inference latency} is better}

Therefore, the goal of this work is to effectively incorporate open-world knowledge while preserving the advantages of classical recommender systems. However, despite the appealing capabilities of LLMs, extracting and utilizing knowledge from them is a non-trivial task. For one thing, LLMs encode vast corpora of world knowledge across various scenarios. Identifying useful external knowledge for recommendation and eliciting the accurate reasoning process on user preferences are quite challenging. Moreover, the world knowledge generated by LLMs is in the form of human-like texts and cannot be interpreted by recommendation algorithms. Even if some LLMs are open-sourced, the decoded outputs are usually large dense vectors (\eg, 4096 for each token), which are highly abstract and not directly compatible with recommender systems. Effectively transforming the output knowledge to be compatible with the recommendation space, without information loss or misinterpretation, is pivotal for the quality of the recommendation. Besides, the knowledge generated by LLMs can sometimes be unreliable or misleading due to the hallucination problem \cite{ji2023survey}. Hence, it is critical to increase the reliability and availability of the generated knowledge to fully unleash the potential of open-world recommender systems.

To address the above problems, we propose an Open-World \underline{K}nowledge \underline{A}ugmented \underline{R}ecommendation Framework with Large Language Models, (dubbed \textit{KAR}). KAR is a model-agnostic framework that bridges classical recommender systems and open-world knowledge, leveraging both reasoning and factual knowledge from the LLMs. By first leveraging LLMs to generate open-world knowledge, and then applying classical recommender systems to model the collaborative signals, we combine the advantages of both LLMs and RSs and significantly improve the model's predictive accuracy. We also propose to prestore the obtained knowledge to avoid the inference latency issue when incorporating LLMs in RSs. 

% \cb{why not use OpenRec directly?}

Specifically, KAR consists of three stages: (1) knowledge reasoning and generation, (2) knowledge adaptation, and (3) knowledge utilization. For knowledge reasoning and generation, to avoid the compositional gap, we propose \textit{factorization prompting} to break down the complex preference reasoning problem into several key factors to generate the reasoning knowledge on users and the factual knowledge on items. Then, the knowledge adaptation stage transforms the generated knowledge to augmented vectors in recommendation space. In this stage, we propose \textit{hybrid-expert adaptor} module to reduce dimensionality and ensemble multiple experts for robust knowledge learning, thus increasing the reliability and availability of the generated knowledge. Finally, in the knowledge utilization stage, the recommendation model incorporates the augmented vectors with original domain features for prediction, combining both the recommendation domain knowledge and the open-world knowledge. Our main contributions can be summarized as follows:
\begin{itemize}
    % \item We present an open-world recommender system exploiting large language models, KAR, which bridges the gap between the recommendation domain knowledge and the open-world knowledge. To the best of our knowledge, this is the first practical solution that introduces logical reasoning with LLMs for user preferences to the recommendation domain.
    % \item KAR transforms the open-world knowledge to dense vectors located in recommendation space, which are compatible with any recommendation models and are flexible to use.
    % \item The knowledge generation and encoding process of KAR can be preprocessed and prestored for fast training and inference, which avoids the large inference latency when using LLMs in RSs.
    \item We present an open-world recommender system, KAR, which bridges the gap between the recommendation domain knowledge and the open-world knowledge from LLMs. To the best of our knowledge, this is the first practical solution that introduces logical reasoning with LLMs for user preferences to the recommendation domain.
    \item KAR transforms the open-world knowledge to dense vectors in recommendation space, which are compatible with any recommendation models. We also release the code of KAR and the generated textual knowledge from LLMs\footnote{Code and knowledge are available at \url{https://gitee.com/mindspore/models/tree/master/research/recommend/KAR} and \url{https://github.com/YunjiaXi/Open-World-Knowledge-Augmented-Recommendation}} to facilitate future research.
    \item The knowledge augmentation can be preprocessed and prestored for fast training and inference, avoiding the large inference latency when adopting LLMs to RSs. Now, KAR has been deployed to Huawei's news and music recommendation platforms and gained a 7\% and 1.7\% improvement in the online A/B test. This is one of the first successful attempts in deploying LLM-based recommender to real-world applications.
\end{itemize}
Extensive experiments conducted on public datasets show that KAR significantly outperforms the state-of-the-art models, and is compatible with various recommendation algorithms. We believe that KAR not only sheds light on a way to inject the knowledge from LLMs into the recommendation models, but also provides a practical framework for open-world recommender systems in large-scale applications.

\section{Related Work}
% This section reviews recent studies on recommendation with pretrained language models (PLMs).
This section reviews studies on recent advances in recommendation with pretrained language models (PLMs).

\subsection{PLM as Recommender Itself}
The emergence of pretrained language models (PLMs) has brought tremendous success in Natural Language Processing (NLP), and PLMs also show great potential in other domains like recommendations~\cite{wu2023survey}. One promising direction is to leverage PLMs as the primary driver of recommendations, allowing PLMs to directly accomplish the recommendation tasks~\cite{li2023large,wu2023survey,yu2023self}.  For example, LMRecSys~\cite{zhang2021language} is one of the earliest attempts to transfer the session-based recommendation task into prompts and evaluate the performance of BERT~\cite{bert} and GPT-2~\cite{gpt2} in movie recommendation. Later, P5~\cite{p5} and M6-Rec~\cite{m6rec} finetune pretrained language models (T5~\cite{t5} or M6~\cite{m6}) by converting multiple recommendation tasks to natural language sequences to incorporate knowledge and semantics inside the training corpora for personalization. Similarly, RecFormer~\cite{li2023text} models user preferences and item features as language representations for sequential recommendation. In this earlier stage, the sizes of the language models for recommendation are relatively small (\eg, under billions of parameters), and finetuning is usually involved for better performance.

With the scaling of the model size and corpus volume, especially with the emergence of ChatGPT~\cite{gpt4}, LLMs have shown uncanny capability in a wide variety of tasks~\cite{fan2023recommender}. One of the unique abilities of LLMs is reasoning, which emerges only when the model size surpasses a certain threshold. Zero-shot learning or in-context learning is widely used since finetuning LLMs requires lots of resources. Several studies apply LLMs as recommenders and achieve some preliminary results~\cite{kang2023llms,liu2023chatgpt,chatrec,nir,dai2023uncovering,hou2023large}. For instance, ChatRec~\cite{chatrec} employs LLMs as a recommender system interface for conversational multi-round recommendations. Liu \etal~\cite{liu2023chatgpt} study whether ChatGPT can serve as a recommender with task-specific prompts and report the zero-shot performance. Hou \etal~\cite{hou2023large} further report the zero-shot ranking performance of LLMs with historical interaction data.

However, directly using LLMs as recommenders generally falls behind state-of-the-art recommendation algorithms, implying the importance of domain knowledge and collaborative signals for recommendation tasks~\cite{kang2023llms,dai2023uncovering,lin2023can}. Therefore, there are also methods exploring the incorporation of recommendation collaborative signals into LLMs through parameter-efficient finetuning approaches~\cite{liu2023pre}. For example, TALLRec~\cite{bao2023tallrec} finetunes LLaMA-7B model~\cite{llama} with a LoRA~\cite{hu2021lora} architecture on recommendation data. Another study~\cite{harte2023leveraging} finetunes an Open-AI ada model\footnote{https://platform.openai.com/docs/guides/fine-tuning} on recommendation data but finds its performance lag behind utilizing the embedding of LLM for similarity matching or as an initialization for recommendation model. 

Moreover, these preliminary studies mainly overlook the inference latency during deployment and the compositional gap problem of LLMs. In this work, we propose factorization prompting to extract both reasoning and factual knowledge from LLMs, alleviating the compositional gap. This knowledge is then adapted to the recommendation domain as augmented representation vectors, which can be prestored for fast training and inference.

% \ljh{What KAR does can be move to the next subsection. Emphasize on the performance and efficiency here briefly is OK.}

% Unlike above approach where PLM is used as an auxiliary component in traditional RS, another category of work leverages PLMs as the primary driver of recommendations, allowing PLMs to generate recommended items~\cite{li2023large,wu2023survey,yu2023self}.

\subsection{PLM as Component of Traditional Recommender}
 Unlike the above approaches where PLMs are used as the primary driver of recommendations, another category of work leverages PLMs as an auxiliary component in traditional RSs. Here, PLMs are usually adopted to encode the textual features (\eg, item descriptions, user reviews) or provide extra knowledge for classical recommendations for better user or item representations~\cite{ubert,zesrec,UniSRec,VQRec}. 
For example, U-BERT~\cite{ubert} leverages the embeddings of user review texts encoded by BERT~\cite{bert} to complement user representations.
ZEREC~\cite{zesrec} incorporates traditional recommender systems with PLMs to generalize from a single training dataset to other unseen testing domains. 
UniSRec~\cite{UniSRec} utilizes BERT~\cite{bert} to encode user behaviors, and therefore learns universal sequence representations for downstream recommendation.
Built upon UniSRec, VQ-Rec~\cite{VQRec} further adopts vector quantization techniques to map language embeddings into discrete codes, balancing the semantic knowledge and domain features.
The above methods usually use small PLMs (\eg, BERT-base with 110M parameters) to convert texts to dense vectors, the semantic information of which could be limited and thus fail to provide strong assistance for traditional recommender systems.

Another line of work adopts large language models  with billion-level parameters, focusing on encoding or prompting external open-world knowledge from LLMs. For instance, some researchers propose S\&R Multi-Domain Foundation model~\cite{gong2023unified}, which finetunes ChatGLM2-6B~\cite{du2022glm} to extract domain invariant features for promoting search and recommendation performance in cold-start scenarios. LLM-Rec~\cite{lyu2023llm} investigates various prompting strategies to generate augmented input text from GPT-3 (\textit{text-davinci-003}), which improves the recommendation capabilities. Another work~\cite{mysore2023large} utilizes InstructGPT (175B)~\cite{ouyang2022training} for authoring synthetic narrative queries from user-item interactions and train retrieval models for narrative-driven recommendation on synthetic data.  TagGPT~\cite{li2023taggpt} provides a system-level solution of tag extraction and multi-modal tagging in a zero-shot fashion equipped with GPT-3.5 (\textit{gpt-3.5-turbo}).
Although these methods have made early attempts at utilizing LLMs, they either simply use LLMs as fixed encoders to convert original texts into dense vectors without generating additional textual knowledge, or do not make specific designs to incorporate the generated textual information into traditional recommender systems. This may lead to the instability of RS due to the noise or high dimension of LLM embeddings.

% \ljh{Need a new paragraph to claim what we have done in this paper. In this paper, we xxxx.}

In this paper, we propose factorization
prompting to break down the complex preference reasoning problem into several key factors to generate open-world knowledge, avoiding the compositional gap. Then, we devise a hybrid-expert adaptor module to reduce dimensionality and ensemble multiple experts for robust knowledge learning, thus increasing the reliability and availability of the generated knowledge. The above two stages can be preprocessed and prestored for fast training and inference, which avoids the large inference latency when using LLMs in RSs. Now, our proposed KAR has been deployed in Huawei news and music recommendation platforms
and has gained significant improvements.

% propose to extract both reasoning and factual knowledge from LLMs to enhance the user and item representations for recommendation models. The open-world knowledge is effectively extracted and adapted to the recommendation domain as augmented representation vectors. Our prestored representations also allow for fast training and inference.

\begin{figure*}
    \centering
    % \vspace{-4pt}
    \includegraphics[width=\textwidth]{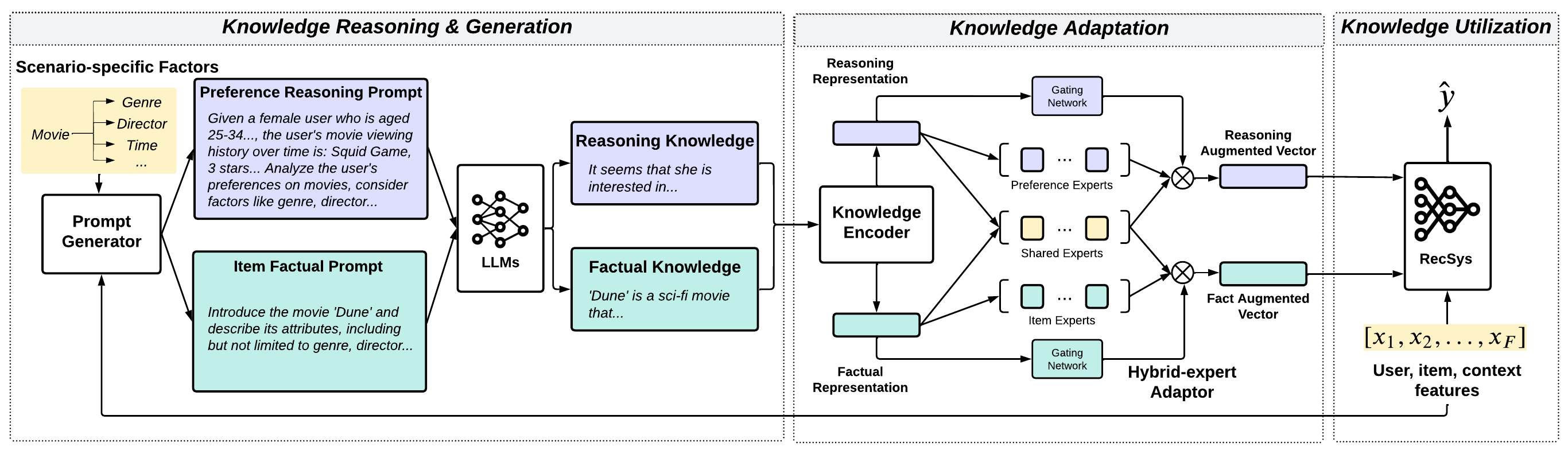}
    % \vspace{-15pt}
    % \caption{The overall framework of KAR, which consists of three stages: (1) Knowledge reasoning and generation; (2) Knowledge adaption; and (3) Knowledge utilization.}
    \caption{The overall framework of KAR, consisting of knowledge reasoning and generation stage, knowledge adaptation stage, and knowledge utilization stage. \textbf{Knowledge reasoning and generation stage} leverages our designed factorization prompting to extract the reasoning and factual knowledge from LLMs. \textbf{Knowledge adaptation stage} converts textual open-world knowledge into compact and the reasoning and fact augmented representations suitable for recommendation. \textbf{Knowledge utilization stage} integrates the reasoning and fact augmented vectors into an existing recommendation model.}
    % \vspace{-5pt}
    \label{fig:open_world}
\end{figure*}
% Click-through rate (CTR) prediction is a crucial task in recommender systems, sharing similar input-output characteristics with many recommendation tasks. In this work, we focus on CTR prediction as the basic task of our framework and this section introduce its formulation and the basic structure. However, it is important to note that our framework is extendable to other recommendation tasks as well.

\section{Preliminaries}\label{sect:network_structure}
In this section, we formulate the recommendation task and introduce the notations. The recommendation task is generally formulated as a binary classification problem over multi-field categorical data. The dataset is denoted as $\mathcal D = \{(x_1, y_1), \ldots, (x_i, y_i), \ldots, (x_n, y_n)\}$, where $x_i$ represents the categorical features for the $i$-th instance and $y_i$ denotes the corresponding binary label (0 for no-click and 1 for click). Usually, $x_i$ contains sparse one-hot vectors from multiple fields, such as item ID and genre. We can represent the feature as $x_i=[x_{i,1}, x_{i,2},\ldots,x_{i, F}]$ with $F$ being the number of field and $x_{i, k}$, $k=1,\ldots,F$ being the feature of the corresponding field. 

Recommendation models usually aims to learn a function $f(\cdot)$ with parameters $\theta$ that can accurately predict the click probability $P(y_i=1|x_i)$ for each sample $x_i$, that is $\hat{y_i} = f(x_i;\theta)$. In practice, industrial recommender systems are confronted with massive users and items. Thus, their recommendation is usually divided into multiple stages, i.e., candidate generation, ranking, and reranking~\cite{liu2022neural}, where different models are used to narrow down the relevant items.
 
However, these classical models are typically trained on a specific recommendation dataset (\ie, a closed system), overlooking the potential benefits of accessing open-world knowledge. 

\section{Methodology}
We first provide an overview of our proposed Open-World Knowledge Augmented Recommendation Framework with Large
Language Models, and then elaborate on the details of each component.

\subsection{Overview}
To extract open-world knowledge from LLMs and incorporate it into RSs, we design KAR, as shown in Figure~\ref{fig:open_world}. This framework is model-agnostic and consists of the following three stages: 

\smallskip
\noindent
\textbf{Knowledge Reasoning and Generation Stage} leverages our designed factorization prompting to extract recommendation-relevant knowledge from LLMs. We first decompose the complex reasoning tasks by identifying major factors that determine user preferences and item characteristics. Then according to each factor, LLMs are required to generate (i) reasoning knowledge on user preferences, and (ii) factual knowledge about items. Thus, we can obtain the open-world knowledge beyond the original recommendation dataset. 

% \Jianghao{This part is quite confusing. I suggest using `itemize` to introduce the two different prompts.}
% The direct knowledge on items and indirect knowledge on user preferences are then generated by LLMs with item and preference prompts, respectively. 

\smallskip
\noindent
\textbf{Knowledge Adaptation Stage} converts textual open-world knowledge into compact and relevant representations suitable for recommendation, bridging the gap between LLMs and RSs. First, the reasoning and factual knowledge obtained from LLMs are encoded into dense representations by a knowledge encoder. Next, a hybrid-expert adaptor is designed to transform the representations from the semantic space\footnote{the embedding space from language models} to the recommendation space. In this way, we obtain the reasoning augmented vector for user preferences and the fact augmented vector for each candidate item.  

% \Jianghao{natural language space? Since we also mention `semantic` in literal CTR prediction papers}

\smallskip
\noindent
\textbf{Knowledge Utilization Stage} integrates the reasoning and fact augmented vectors into an existing recommendation model, enabling it to leverage both domain knowledge and open-world knowledge during the recommendation process. 

The knowledge generation and encoding are conducted through preprocessing. The hybrid-expert adaptor and recommendation model are jointly trained in an end-to-end manner.

\subsection{Knowledge Reasoning and Generation}
As the model size scales up, LLMs can encode a vast array of world knowledge and have shown emergent behaviors such as the reasoning ability~\cite{qiao2023reasoning,huang2022reasoning}. This opens up new possibilities for incorporating reasoning knowledge for user preferences and factual knowledge for candidate items in recommendation systems. However, it is non-trivial to extract the reasoning knowledge and corresponding factual knowledge from LLMs due to the following two challenges.

Considering the reasoning knowledge, according to~\cite{press2022measuring}, LLMs often suffer from the \textit{compositional gap} where the model fails at generating the correct answer to the compositional question but can correctly answer all its sub-questions. User's clicks on items are motivated by multiple key aspects and user's interests are diverse and multifaceted, which involve multiple reasoning steps. To this end, LLMs may not be able to directly produce accurate reasoning knowledge. Expecting LLMs to provide precise recommendations in one step as in previous work~\cite{liu2023chatgpt,chatrec,hou2023large} might be overly ambitious. 
% By breaking down the complex preference reasoning problem into simpler subproblems of reasoning according to each individual factor, the compositional problem can be alleviated and correlations of different factors behind the user preferences can be analyzed.

As for the factual knowledge, LLMs contain massive world knowledge, yet not all of it is useful for recommendation. When the request to an LLM is too general, the generated factual knowledge may be correct but useless, as it may not align with the inferred user preferences. For example, an LLM may infer that a user may prefer highly acclaimed movies that have received multiple awards, while the generated factual knowledge is about the storyline of the target movie. This mismatch between preference reasoning knowledge and item factual knowledge may limit the performance of RSs.  

% As for the factual knowledge, LLMs contain massive world knowledge, yet not all of them are useful for recommendation. By augmenting the items with the preference-sensitive feature factors, the recommender systems can easily match the attracting items to the users. For example, the LLM infers that a user may prefer highly acclaimed movies that have received multiple awards, based on the user's history and the external knowledge it possesses. However, in a movie recommendation dataset, the awards of the movies may not be available. Incorporating the award knowledge from the LLMs can help align the items to the user preferences, enables more accurate and personalized recommendations. 

Therefore, inspired by the success of Factorization Machines \cite{rendle2010factorization} in RSs, we design \textit{factorization prompting} to explicitly "factorize" user preferences into several major factors for effectively extracting the open-world knowledge from LLMs. With the factors incorporated into \textit{preference reasoning prompt}, the complex preference reasoning problem can be broken down into simpler subproblems for each factor, thus alleviating the compositional gap of LLMs. Besides, we also design \textit{item factual prompt} which utilizes those factors to extract factual knowledge relevant to user preferences. This ensures the generated reasoning knowledge and factual knowledge are aligned for the effective utilization in RSs.

\begin{figure*}
    \centering
    \vspace{-4pt}
    \includegraphics[width=0.99\textwidth]{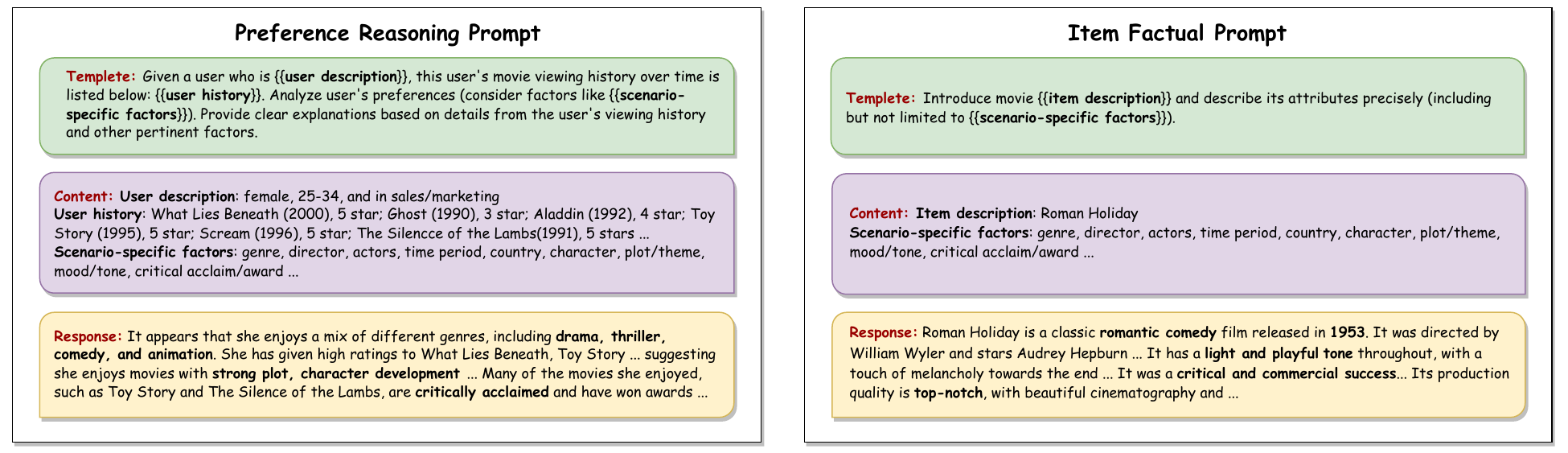}
    \vspace{-3pt}
    \caption{Example prompts for KAR. The green, purple, and yellow text bubbles represent the prompt template, the content to be filled in the template, and the response generated by LLMs, respectively (some text has been omitted due to the page limits).}
    \vspace{-3pt}
    \label{fig:promp}
\end{figure*}

\vspace{-3pt}
\subsubsection{Scenario-specific Factors} 

% \Jianghao{... solve complex reasoning tasks that could be reduced to a chain of thought steps}
% Therefore, it is necessary to augment the candidate items with factual knowledge aligned with user preferences, such as their awards information.
 % User preferences are inherently complicated and multifaceted,

The factors determining user preferences may vary for different recommendation scenarios. To determine the factors for different scenarios, we rely on a combination of interactive collaboration with LLMs and expert opinions. For example, in movie recommendation, given a prompt "\textit{List the important factors or features that determine whether a user will be interested in a movie}", LLMs can provide some potential factors. Then, we involve human experts to confirm and refine the outputs to acquire the final scenario-specific factors for movie recommendation --- including \textit{genre, actors, directors, theme, mood, production quality, and critical acclaim}. Similarly, in news recommendation, we may obtain the factors like \textit{topic, source, region, style, freshness, clarity, and impact}. This collaborative process between LLMs and experts ensures that the chosen factors encompass the critical dimensions of user preference and item characteristics for each scenario. The specification of these factors is required only once for each scenario, showing that the proposed method can be easily generalized to different scenarios with little human intervention.
% \xx{While there is human intervention involved here, it is required only once for each scenario. Compared to the extensive human annotation required for knowledge graph, the human involvement in this case is minimal and relatively straightforward.}

% The scenario-specific factors will be leveraged in following prompt designed for user preference reasoning and item factual knowledge generation. On one hand, incorporating scenario-specific factors into the preference reasoning prompt can alleviate the compositional gap. LLMs only need infer user preferences regarding to those factors separately, making preference reasoning task more manageable for LLMs. On the other hand, including the same factors in the prompt for item factual knowledge generation enable LLMs to generate aligned knowledge that is relevant to user preference. This ensure the generated knowledge can be effectively utilized by the RSs

% This not only allows LLMs to solve the preference reasoning problem more easily with less compositional gap and 
% By breaking down user preference, we allow LLMs to recall more
% detailed knowledge and solve the preference reasoning problem
% more easily, thereby alleviating the compositional gap problem and fully leveraging the reasoning capabilities of LLMs

% \Jianghao{This section only talks about preference reasoning prompt?}

% Such complex reasoning ability is usually unlocked by specific prompting strategies, e.g., generated knowledge prompting~\cite{}, and problem decomposition ~\cite{}. 
\subsubsection{LLM as Preference Reasoner \& Knowledge Provider}
After obtaining the scenario-specific factors, we introduce them into our prompt engineering. To extract reasoning and factual knowledge from the open world, we propose to apply the LLM as a \textit{preference reasoner} to infer user preferences, and a \textit{knowledge provider} to acquire external factual knowledge for candidate items. Therefore, we design two types of prompts accordingly: preference reasoning prompt and item factual prompt, as illustrated in Figure~\ref{fig:promp}.

\textbf{Preference reasoning prompt} is constructed with the user's profile description, behavior history, and scenario-specific factors. Figure~\ref{fig:promp} shows an example of the prompt and real response from the LLM, where the user profile description and behavior history provide LLM with the necessary context and user-specific information to understand the user's preferences. Scenario-specific factors can instruct the LLM to analyze user preference from different facets and allow the LLM to recall the relevant knowledge more effectively and comprehensively. For example, in the factor of genre, the LLM infers user preference for genres such as thriller, comedy, and animation based on user's positive ratings on thriller movies like \textit{What Lies Beneath}, \textit{Scream}, as well as comedy animations like \textit{Toy Story} and \textit{Aladdin}. With the designed prompt, LLM can successfully analyze the user's preferences toward corresponding factors, which is beneficial for recommendations.
 % We can observe that in the generated response, the LLM analyzes and identifies the user's preferences for the corresponding factors based on the user's profile and behavior history, which is beneficial for recommendations.

% Instructing the LLM to analyze the user preference according to the scenario-specific factors allows the LLM to recall the relevant knowledge more effectively and comprehensively.

\textbf{Item factual prompt} is designed to fill the knowledge gap between the candidate items and the generated reasoning knowledge. Since the dataset in RS may lack relevant knowledge about scenario-specific factors from items, we need to extract corresponding knowledge from LLM to align the generated user and item knowledge.  As illustrated in Figure~\ref{fig:promp}, an item prompt consists of two parts -- the target item description and the scenario-specific factors. This prompt can guide LLM in compensating for the missing knowledge within the dataset. For instance, in Figure~\ref{fig:promp}, LLM supplements \textit{Roman Holiday} with \textit{"a light and playful tone"}, which is rarely recorded in the datasets. In this way, LLM provides external knowledge that aligns with user preferences, allowing for more accurate and personalized recommendations.

By combining the two kinds of prompts, we enable the LLM to act as both a preference reasoner and a knowledge provider, thereby extracting the open-world knowledge from LLMs.
 % and expanding the knowledge scope of RSs

% We refer to the generated textual response with the preference and item prompt as the \textit{preference knowledge} and the \textit{item knowledge}, respectively. 
\vspace{-5pt}
\subsection{Knowledge Adaptation}
The knowledge generated by LLMs presents new challenges in harnessing its potential to assist recommendation models: 1) The knowledge generated by LLMs is usually in the form of text, which cannot be directly leveraged by traditional RSs that typically process categorical features. 2) Even if some LLMs are open-sourced, the decoded outputs are usually large dense vectors (\eg, 4096 for each token) and lie in a semantic space that differs significantly from the recommendation space. 3) The generated knowledge may contain noise or unreliable information~\cite{ji2023survey}. 

To address these challenges, we have devised two modules: a \textit{knowledge encoder} and a \textit{hybrid-expert adaptor}. The knowledge encoder module encodes the generated textual knowledge into dense vectors and aggregates them effectively. The hybrid-expert adaptor converts dense vectors from the semantic space to the recommendation space. It tackles dimensionality mismatching and allows for noise mitigation. Thus, the knowledge adaptation stage increases the reliability and availability of the generated knowledge and bridges the gap between LLMs and RSs.
\subsubsection{Knowledge Encoder} To harness the potential of textual knowledge generated by LLMs, we employ a knowledge encoder, \eg, BERT~\cite{bert}, to obtain the encodings for each token within the text. Then, we require an aggregation process that combines each token to generate the \textit{preference reasoning representation} $r^p_i\in\mathbb{R}^{m}$ and the \textit{item factual representation} $r^\iota_i\in\mathbb{R}^{m}$ of size $m$ as follows
\begin{equation}
    \begin{split}       
    &r^p_i = \text{Aggr}(\text{Encoder}(klg^p_i))\,, \\
    &r^\iota_i = \text{Aggr}(\text{Encoder}(klg^\iota_i))\,,
    \end{split}
\label{eq:aggr_encoder}
\end{equation}
where $klg^p_i$ and $klg^\iota_i$ denote the textual reasoning knowledge and factual knowledge generated by LLMs of the $i$-th instance in the dataset. Here, various aggregation functions can be employed, such as the representation of the [CLS] token and average pooling. In practice, we primarily adopt average pooling. Note that the knowledge encoder is devised for situations where we only have access to the textual outputs of LLMs. If the dense vector outputs from LLMs are available, a separate knowledge encoder can be eliminated.
% In practice, we primarily adopt average pooling, but we also explore the effectiveness of other aggregation methods.

% the representation of the last token \Jianghao{Suggest remove `the representation of the last token`; cls and average pooling are OK.},

\subsubsection{Hybrid-expert Adaptor} To effectively transform and compact the attained aggregated representations from the semantic space to the recommendation space, we propose a hybrid-expert adaptor module. The aggregated representations capture diverse knowledge from multiple aspects, so we employ a structure that mixes shared and dedicated experts, inspired by the Mixture of Experts (MoE)~\cite{MoE} approach. This allows us to fuse knowledge from different facets and benefits from the inherent robustness offered by multiple experts. 
% \cb{why not use OpenRec directly?}

In particular, to fully exploit the shared information of the preference reasoning representation and the item factual representation, we have designed both shared experts and dedicated experts for each kind of representation. 
% \cb{formalized into two kinds of adaptation task?}
The shared experts capture the common aspects, such as shared features, patterns, or concepts, that are relevant to both preference reasoning and item factual knowledge. Reasoning and factual representations also have their dedicated sets of experts to capture the unique characteristics specific to the reasoning or factual knowledge. 
Mathematically, denote $\mathcal S_s$, $\mathcal S_p$, and $\mathcal S_\iota$ as the sets of shared experts and dedicated experts for preference reasoning and item factual knowledge with the expert number of $n_s$, $n_p$ and $n_\iota$. The output is the \textit{reasoning augmented vector} $\hat{r}_{i}^p\in\mathbb{R}^{q}$ and the \textit{fact augmented vector} $\hat{r}_{i}^\iota\in\mathbb{R}^{q}$ of size $q$ ($q$ is much less than the original dimension $m$), which are calculated as follows
\begin{equation}
\begin{split}
    \alpha^p_{i} &=\operatorname{Softmax}(g^p(r^p_i))\,,\quad\alpha_{i}^\iota =\operatorname{Softmax}(g^\iota(r^\iota_i))\,,\\
    \hat{r}_{i}^p&=\sum\nolimits_{e\in \mathcal S_s} \alpha_{i,e}^p \times e(r^p_i) + \sum\nolimits_{e\in \mathcal S_p} \alpha_{i,e}^p \times e(r^p_i)\,,\\
    \hat{r}_{i}^\iota&=\sum\nolimits_{e\in \mathcal S_s} \alpha_{i,e}^\iota \times e(r^\iota_i) + \sum\nolimits_{e\in \mathcal S_\iota} \alpha_{i,e}^\iota \times e(r^\iota_i)\,,\\
\end{split}
\end{equation}
where $g^p(\cdot)$ and $g^\iota(\cdot)$ are the gating networks for preference reasoning and item factual representations, and their outputs $\alpha^p_i$ and $\alpha^\iota_i$ are of size $n_s+n_p$ and $n_s+n_\iota$. Here $e(\cdot)$ denotes the expert network, and $\alpha_{i,e}^p$ and $\alpha_{i,e}^\iota$ are the weights of expert $e(\cdot)$ generated by the gating network for preference and item, respectively. Here, each expert network $e(\cdot)$ is designed as Multi-Layer Perceptron (MLP), facilitating dimensionality reduction and space transformation.
% \cb{the optimization of this stage is missing}

\subsection{Knowledge Utilization}
Once we have obtained the reasoning augmented vector and the fact augmented vector, we can then incorporate them into backbone recommendation models. In this section, we explore a straightforward approach where these augmented vectors are directly treated as \textbf{additional input features}. Specifically, we use them as additional feature fields in recommendation models, allowing them to explicitly interact with other features. During training, the hybrid-expert adaptor module is jointly optimized with the backbone model to ensure that the transformation process adapts to the current data distribution. Generally, KAR can be formulated as
\begin{equation}
    \hat{y_i} = f(x_i,h_i,\hat{r}_{i}^p,\hat{r}_{i}^\iota;\theta)\,,
     \label{eq:ctr-goal-3}
\end{equation}
which is enhanced by the the reasoning augmented vector $\hat{r}_{i}^p$ and the fact augmented vector $\hat{r}_{i}^\iota$. Importantly, KAR only modifies the input of the backbone model and is independent of the design and loss function of backbone model, so it is flexible and compatible with various backbone model designs. Furthermore, it can be extended to various recommendation tasks, such as sequential recommendation and direct recommendation, by simply adding two augmented vectors in the input. By incorporating the knowledge augmented vectors, KAR combines both the open-world knowledge and the recommendation domain knowledge in a unified manner to provide more informed and personalized recommendations.

% Importantly, our proposed KAR can be applied for different backbone models and loss functions, demonstrating its flexibility and compatibility.

\vspace{-6pt}
\subsection{Speed-up Approach}\label{sec:speed_up}
Our proposed KAR framework adopts LLMs to generate reasoning knowledge for user preferences and factual knowledge for candidate items. Due to the immense scale of the model parameters, the inference of LLMs takes extensive computation time and resources, and the inference time may not meet the latency requirement in real-world recommender systems with large user and item sets. 

To address this, we employ an acceleration strategy to \textbf{prestore knowledge representations} $r^p_i$ and $r^\iota_i$ generated by the knowledge encoder or the LLM into a database. As such, we only use the LLM and knowledge encoder once before the training of backbone models. During the training and inference of the backbone model, relevant representations are retrieved from the database. Besides, the efficiency of offline knowledge generation with LLM can further be enhanced via quantization or hardware acceleration techniques.
% \Jianghao{Suggest: During the training and inference stages, according to the user ID and target item ID, knowledge representations are retrieved from the database for the backbone model.} 
% \Jianghao{Suggest new paragraph here. Next is other further speedup for inference.}
% \Jianghao{$r^p_i$ and $r^\iota_i$ are generated by knowledge encoder, instead of LLM. Suggest: prestore the knowledge representation $r^p_i$ and $r^\iota_i$ generated by the knowledge encoder in the adaptation stage into a database}

If we have stricter requirements for inference time or storage efficiency, we can detach the adaptor from the model after training and further \textbf{prestore augmented vectors} \ie, $\hat{r}_{i}^p$ and $\hat{r}_{i}^\iota$, for inference. The dimension of the augmented vectors (\eg, 32) is usually much smaller than that of the knowledge representations (\eg, 4096), which improves the storage efficiency. Additionally, prestoring the augmented vectors reduces the inference time to nearly the same as the original backbone model, and we have provided experimental verification in Section~\ref{sec:efficiency}. In particular, assume the inference time complexity of the backbone model is $O(f(n, m))$, where $n$ is the number of fields and $m$ is the embedding size. The polynomial function $f(n,m)$ varies depending on different backbone models. With KAR, the inference time complexity is $O(f(n+2,m))=O(f(n,m))$, which is equivalent to the complexity of the original model. 

% In particular, assume the inference time complexity of backbone model is $O(f(n))$, where $n$ is the number of fields and $f(n)$ varies depending on different backbone models. In our framework, the inference time complexity is $O(f(n+2))=O(f(n))$, which is equivalent to the complexity of original model. 
% \Jianghao{Complexity should be $O(f(n,m))$, where $n$ is the number of fields and $m$ is the embedding size.}
% \Jianghao{$f(n,m)$ is a polynomial function, which varies depending on different backbone models}

Since item features are relatively fixed and do not change frequently, it is natural and feasible to prestore the item factual knowledge for further use. Moreover, user behaviors evolve over time, making it challenging for LLMs to provide real-time reasoning knowledge about behaviors. However, considering that long-term user preferences are relatively stable, and the backbone model already emphasizes modeling recent user behaviors, it is unnecessary to require LLMs to have access to real-time behaviors. Therefore, LLM can infer long-term preferences based on users' long-term behaviors, allowing for conveniently prestoring the generated knowledge without frequent updates. The inference overhead of LLMs can also be significantly reduced. The backbone models can capture ever-changing short-term preferences with timely model updates.  This can take better advantage of both LLMs and recommendation models. Similar to common practice for cold start users or items~\cite{zhang2014addressing}, we use default vectors when encountering new users or items at inference time. Subsequently, we will generate the knowledge for those new users and items offline and add them to the database.

\vspace{-5pt}
\section{Experiment}
% Click-through rate (CTR) prediction is a crucial task in recommender systems, sharing similar input-output formualtion with many recommendation tasks. Thus, this section focuses on evaluating the effectiveness of our proposed framework, KAR, with CTR prediction as the fundamental task. Importantly, KAR is easily extendable to other recommendation tasks. 
To gain more insights into KAR, we tend to address the following research questions (RQs) in this section. 
\begin{itemize}
    \item \textbf{RQ1:} What improvements can KAR bring to backbone models on different tasks, such as CTR prediction and reranking? 
    \item \textbf{RQ2:} How does KAR perform compared with other PLM-based baseline methods?
    \item \textbf{RQ3:} Does the knowledge from LLM outperform other methods of knowledge, such as knowledge graph?
    \item \textbf{RQ4:}Does KAR gain performance improvement when deployed online?
    \item \textbf{RQ5:} How do the reasoning knowledge and factual knowledge generated by the LLM contribute to performance improvement?
    \item \textbf{RQ6:} How do different knowledge adaptation approaches impact the performance of KAR?
    \item \textbf{RQ7:} Does the acceleration strategy, preprocessing and prestorage, enhance the inference speed?
\end{itemize}
By answering these questions, we aim to comprehensively evaluate the performance and versatility of our proposed framework.
% What roles do the preference and item knowledge generated by the LLM play?
\vspace{-4pt}
\subsection{Setup}
% \footnote{Considering the large API cost of the LLMs, we only conduct experiments on one public dataset.}
\subsubsection{Dataset}
Our experiments are conducted on public datasets, MovieLens-1M\footnote{\url{https://grouplens.org/datasets/movielens/1m/}} and Amazon\footnote{\url{https://cseweb.ucsd.edu/~jmcauley/datasets/amazon_v2/}}. \textbf{MovieLens-1M} contains 1 million ratings provided by 6000 users for 4000 movies. Following the data processing similar to DIN~\cite{DIN}, we convert the ratings into binary labels by labeling ratings of 4 and 5 as positive and the rest as negative. The data is split into training and testing sets based on user IDs, with 90\% assigned to the training set and 10\% to the testing set. The dataset contains user features like age, gender, occupation, and item features like item ID and category. The input to the models are user features, user behavior history (the sequence of viewed movies with their ID, category, and corresponding ratings), and target item features. \textbf{Amazon-Book}~\cite{ni2019justifying} is the “Books” category of the Amazon Review Dataset. After filtering out the less-interacted users and items, we remain 11, 906 users and 17, 332 items with 1, 406, 582 interactions. The preprocessing is similar to MovieLens-1M, with the difference being the absence of user features. Additionally, ratings of 5 are regarded as positive and the rest as negative.

% with 51, 311, 621 reviews and 2, 935, 525 books. Due to the large amount of data, we filter out users with fewer than 60 interactions and items with fewer than 40 interactions.

% covering multiple categories. Here we mainly focus on category \textbf{Books}
% \footnote{To avoid the large API cost, we select one of the most commonly used dataset for recommender systems.}
% The features include the ID and category of the target movie, profile features of the user, as well as the ID, category, and rating of the movies that the user has previously viewed. Our goal is to predict whether a user will rate a given movie above 3 (positive label) based on features of the movie and user's historical behaviors. 
\subsubsection{Backbone Models} Because KAR is a model-agnostic framework, various tasks and models in traditional ID-based recommendation can serve as the backbone model, taking as input the knowledge-augmented vectors generated by KAR. Here, we select two crucial recommendation tasks: \textbf{CTR prediction} and \textbf{reranking}, to validate the effectiveness of KAR across various tasks. CTR prediction aims to anticipate how likely a user is to click on an item, usually used in the ranking stage of recommendation. Reranking is to reorder the items from the previous ranking stage and derive a list that yields more utility and user satisfaction~\cite{liu2022neural}.

We choose 9 representative CTR models as our backbone models, which can be categorized into  user behavior models and feature interaction models. \textbf{User Behavior Models} emphasize modeling sequential dependencies of user behaviors. \textbf{DIN}~\cite{DIN} utilizes attention to model user interests dynamically with respect to a certain item. \textbf{DIEN}~\cite{DIEN} extends DIN by introducing an interest evolving mechanism to capture the dynamic evolution of user interests over time. 

% The former focuses on modeling feature interactions between different feature fields and we adopt widely-used \textbf{DeepFM}~\cite{DeepFM}, \textbf{xDeepFM} \cite{xDeepFM}, \textbf{DCN}~\cite{DCN}, \textbf{DCNv2}~\cite{DCNv2}, \textbf{FiBiNet}~\cite{FiBiNET}, \textbf{FiGNN}~\cite{Fi-GNN}, and \textbf{AutoInt}~\cite{AutoInt}. The latter emphasizes on modeling sequential dependencies of user behaviors and we employ \textbf{DIN}~\cite{DIN} and \textbf{DIEN}~\cite{DIEN}. 

\begin{table*}[h]
\centering
% \vspace{-4pt}
\caption{Comparison between KAR and backbone CTR prediction models.}
% \vspace{-4pt}
\scalebox{1.06}{
\setlength{\tabcolsep}{1.3mm}{
\begin{tabular}{ccccccc|cccccc}
\toprule
\multirow{3}{*}{\textbf{\begin{tabular}[c]{@{}c@{}}Backbone\\ model\end{tabular}}} & \multicolumn{6}{c|}{\textbf{MovieLens-1M}} & \multicolumn{6}{c}{\textbf{Amazon-Books}} \\
\cmidrule{2-13}
 & \multicolumn{3}{c}{\textbf{AUC}} & \multicolumn{3}{c|}{\textbf{LogLoss}} & \multicolumn{3}{c}{\textbf{AUC}} & \multicolumn{3}{c}{\textbf{LogLoss}} \\
 \cmidrule{2-13}
 & \textbf{base} & \textbf{KAR} & \textbf{Improv.} & \textbf{base} & \textbf{KAR} & \textbf{Improv.} & \textbf{base} & \textbf{KAR} & \textbf{Improv.} & \textbf{base} & \textbf{KAR} & \textbf{Improv.}  \\
 \midrule
% DCNv2 & 0.7924 & \textbf{0.8049*} & 1.58\% & 0.5451 & \textbf{0.5315*} & 2.50\% & 0.8269 & \textbf{0.8350*} & 0.98\% & 0.4973 & \textbf{0.4865*} & 2.17\% \\
% DCNv1 & 0.7929 & \textbf{0.8044*} & 1.46\% & 0.5457 & \textbf{0.5319*} & 2.53\% & 0.8268 & \textbf{0.8348*} & 0.97\% & 0.4973 & \textbf{0.4869*} & 2.11\% \\
% DeepFM & 0.7928 & \textbf{0.8041*} & 1.44\% & 0.5462 & \textbf{0.5321*} & 2.57\% & 0.8269 & \textbf{0.8347*} & 0.94\% & 0.4969 & \textbf{0.4873*} & 1.93\% \\
% FiBiNet & 0.7925 & \textbf{0.8051*} & 1.59\% & 0.5450 & \textbf{0.5310*} & 2.56\% & 0.8269 & \textbf{0.8351*} & 0.99\% & 0.4973 & \textbf{0.4870*} & 2.07\% \\
% AutoInt & 0.7934 & \textbf{0.8060*} & 1.59\% & 0.5440 & \textbf{0.5297*} & 2.65\% & 0.8262 & \textbf{0.8357*} & 1.16\% & 0.4981 & \textbf{0.4863*} & 2.37\% \\
% FiGNN & 0.7944 & \textbf{0.8054*} & 1.39\% & 0.5424 & \textbf{0.5307*} & 2.16\% & 0.8270 & \textbf{0.8352*} & 0.99\% & 0.4977 & \textbf{0.4870*} & 2.14\% \\
% xDeepFM & 0.7942 & \textbf{0.8041*} & 1.25\% & 0.5457 & \textbf{0.5317*} & 2.57\% & 0.8271 & \textbf{0.8351*} & 0.97\% & 0.4971 & \textbf{0.4866*} & 2.10\% \\
% DIEN & 0.7960 & \textbf{0.8059*} & 1.25\% & 0.5469 & \textbf{0.5298*} & 3.13\% & 0.8307 & \textbf{0.8391*} & 1.01\% & 0.4926 & \textbf{0.4812*} & 2.32\% \\
% DIN & 0.7975 & \textbf{0.8066*} & 1.15\% & 0.5387 & \textbf{0.5304*} & 1.55\% & 0.8304 & \textbf{0.8418*} & 1.38\% & 0.4937 & \textbf{0.4801*} & 2.77\%\\
DCNv2                                                                              & 0.7830          & \textbf{0.7935*} & 1.34\%           & 0.5516          & \textbf{0.5410*} & 1.92\%           & 0.8269        & \textbf{0.8350*} & 0.98\%           & 0.4973        & \textbf{0.4865*} & 2.17\%           \\
DCNv1                                                                              & 0.7828          & \textbf{0.7927*} & 1.28\%           & 0.5528          & \textbf{0.5411*} & 2.12\%           & 0.8268        & \textbf{0.8348*} & 0.97\%           & 0.4973        & \textbf{0.4869*} & 2.11\%           \\
DeepFM                                                                             & 0.7824 & \textbf{0.7919*} & 1.22\%           & 0.5518 & \textbf{0.5432*} & 1.56\%  & 0.8269        & \textbf{0.8347*} & 0.94\%           & 0.4969        & \textbf{0.4873*} & 1.93\%           \\
FiBiNet                                                                            & 0.7820          & \textbf{0.7936*} & 1.49\%           & 0.5531          & \textbf{0.5405*} & 2.27\%           & 0.8269        & \textbf{0.8351*} & 0.99\%           & 0.4973        & \textbf{0.4870*} & 2.07\%           \\
AutoInt                                                                            & 0.7821          & \textbf{0.7931*} & 1.40\%           & 0.5520          & \textbf{0.5430*} & 1.62\%           & 0.8262        & \textbf{0.8357*} & 1.16\%           & 0.4981        & \textbf{0.4863*} & 2.37\%           \\
FiGNN                                                                              & 0.7832          & \textbf{0.7935*} & 1.32\%           & 0.5510          & \textbf{0.5437*} & 1.33\%           & 0.8270        & \textbf{0.8352*} & 0.99\%           & 0.4977        & \textbf{0.4870*} & 2.14\%           \\
xDeepFM                                                                            & 0.7823          & \textbf{0.7926*} & 1.32\%           & 0.5520          & \textbf{0.5420*} & 1.81\%           & 0.8271        & \textbf{0.8351*} & 0.97\%           & 0.4971        & \textbf{0.4866*} & 2.10\%           \\
DIEN                                                                               & 0.7853          & \textbf{0.7953*} & 1.27\%           & 0.5494          & \textbf{0.5394*} & 1.83\%           & 0.8307        & \textbf{0.8391*} & 1.01\%           & 0.4926        & \textbf{0.4812*} & 2.32\%           \\
DIN                                                                                & 0.7863          & \textbf{0.7961*} & 1.24\%           & 0.5486          & \textbf{0.5370*} & 2.10\%           & 0.8304        & \textbf{0.8418*} & 1.38\%           & 0.4937        & \textbf{0.4801*} & 2.77\%          \\
\bottomrule
\end{tabular}
}
}
\footnotesize \flushleft\hspace{0.0cm} $*$ denotes statistically significant improvement over backbone CTR prediction models (t-test with $p$-value $<$ 0.05).
% \vspace{-4pt}
\label{tab:backbone_model}
\end{table*}
\textbf{Feature Interaction Models} focus on modeling feature interactions between different feature fields. \textbf{DeepFM}~\cite{DeepFM} is a classic CTR model that combines factorization machine (FM) and neural network to capture low-order and high-order feature interactions. \textbf{xDeepFM} \cite{xDeepFM} leverages the power of both deep network and Compressed Interaction Network to generate feature interactions at the vector-wise level. \textbf{DCN}~\cite{DCN} incorporates cross-network architecture and the DNN model to learn the bounded-degree feature interactions. \textbf{DCNv2}~\cite{DCNv2} is an improved framework of DCN which is more practical in large-scale industrial settings. \textbf{FiBiNet}~\cite{FiBiNET} can dynamically learn the feature importance by Squeeze-Excitation network and fine-grained feature interactions by bilinear function. \textbf{FiGNN}~\cite{Fi-GNN} converts feature interactions into modeling node interactions on the graph for modeling feature interactions in an explicit way. \textbf{AutoInt}~\cite{AutoInt} adopts a self-attentive neural network with residual connections to model the feature interactions explicitly.

As fo reranking task, we implement the state-of-the-art models, \eg, \textbf{DLCM}~\cite{ai2018learning}, \textbf{PRM}~\cite{pei2019personalized}, \textbf{SetRank}~\cite{pang2020setrank}, and \textbf{MIR}~\cite{xi2022multi}, as backbone models. \textbf{DLCM} ~\cite{ai2018learning} first applies GRU to encode and rerank the top results. \textbf{PRM}~\cite{pei2019personalized} employs self-attention to model the mutual influence between any pair of items and users' preferences. \textbf{SetRank}~\cite{pang2020setrank} learns permutation-equivariant representations for the inputted items via self-attention. \textbf{MIR} \cite{xi2022multi} models the set-to-list interactions between candidate set and history list with personalized long-short term interests.

\subsubsection{PLM-based Baselines}
As for baselines, we compare KAR with methods that leverage pretrained language model to enhance recommendation, such as P5~\cite{p5}, UniSRec~\cite{UniSRec}, VQRec~\cite{VQRec}, TALLRec~\cite{bao2023tallrec}, and LLM2DIN~\cite{harte2023leveraging}. 
\textbf{P5}~\cite{p5} is a text-to-text paradigm that unifies recommendation tasks and learns different tasks with the same language modeling objective during pretraining. \textbf{UniSRec}~\cite{UniSRec} designs a universal sequence representation learning approach for sequential recommenders, which introduces contrastive pretraining tasks to effective transfer across scenarios. 
\textbf{VQ-Rec}~\cite{VQRec} uses Vector-Quantized item representations and a text-to-code-to-representation scheme, achieving effective cross-domain and cross-platform sequential recommendation. \textbf{TALLRec}~\cite{bao2023tallrec} finetunes LLaMa-7B~\cite{llama} with a LoRA architecture on recommendation tasks and enhances the recommendation capabilities of LLMs in few-shot scenarios. In our experiment, we implement TALLRec with LLaMa-2-7B-chat\footnote{https://huggingface.co/meta-llama/Llama-2-7b-chat-hf}, since it has better performance and ability of instruction following. According to~\cite{harte2023leveraging}, we design \textbf{LLM2DIN} which initializes DIN with item
embeddings obtained from chatGLM\cite{du2022glm}. We utilize the publicly available code of these three models and adapt the model to the CTR task with necessary minor modifications. We also align the data and features for all the methods to ensure fair comparisons.

\subsubsection{Evaluation Metrics} On CTR prediction task, we employ widely-used \textit{AUC} (Area under the ROC curve) and \textit{LogLoss} (binary cross-entropy loss) as evaluation metrics following~\cite{DCNv2,AutoInt,DeepFM,DIN}. A higher AUC value or a lower Logloss value, even by a small margin (\eg, 0.001), can be viewed as a significant improvement in CTR prediction performance, as indicated by previous studies~\cite{xDeepFM,DCNv2}. As for reranking task, several widely used metrics, \textit{NDCG@K}~\cite{ndcg} and \textit{MAP@K}~\cite{yue2007support}, are adopted, following previous work~\cite{xi2022multi,ai2018learning,pei2019personalized}.

\subsubsection{Implementation Details} 
% We utilize ChatGPT\footnote{\url{https://openai.com/product/chatgpt}} (gpt-3.5-turbo) from OpenAI as the LLM to generate reasoning and factual knowledge. 
We utilize API of a widely-used LLM for generating reasoning and factual knowledge. Then, ChatGLM~\cite{du2022glm} is employed to encode the knowledge, followed by average pooling as the aggregation function in Eq.~\eqref{eq:aggr_encoder}. Each expert in the hybrid-expert adaptor is implemented as an MLP with a hidden layer size of [128, 32]. The number of experts varies slightly across different backbone models, typically with 2-5 shared experts and 2-6 dedicated experts. We keep the embedding size of the backbone model as 32, and the output layer MLP size as [200, 80]. Other parameters, such as batch size and learning rate, are determined through grid search to achieve the best results. For fair comparisons, the parameters of the backbone model and the baselines are also tuned to achieve their optimal performance.
\begin{table*}[h]
\centering
% \vspace{-5pt}
\caption{The comparison of KAR and backbone reranking models on Amazon-Books dataset.}
% \vspace{-5pt}
\scalebox{1.05}{
\setlength{\tabcolsep}{1.3mm}{
\begin{tabular}{ccccccc|cccccc}
\toprule
\multirow{2}{*}{\textbf{\begin{tabular}[c]{@{}c@{}}Backbone\\ Model\end{tabular}}} & \multicolumn{3}{c}{\textbf{MAP@3}} & \multicolumn{3}{c|}{\textbf{MAP@7}} & \multicolumn{3}{c}
{\textbf{NDCG@3}} & \multicolumn{3}{c}{\textbf{NDCG@7}} \\
\cmidrule{2-13}
 & \textbf{base} & \textbf{KAR} & \textbf{Improv.} & \textbf{base} & \textbf{KAR} & \textbf{Improv.} & \textbf{base} & \textbf{KAR} & \textbf{Improv.} & \textbf{base} & \textbf{KAR} & \textbf{Improv.} \\
 \midrule
DLCM & 0.6365 & \textbf{0.6654*} & 4.54\% & 0.6247 & \textbf{0.6512*} & 4.24\% & 0.5755 & \textbf{0.6109*} & 6.15\% & 0.6891 & \textbf{0.7142*} & 3.64\% \\
PRM & 0.6488 & \textbf{0.6877*} & 6.00\% & 0.6359 & \textbf{0.6722*} & 5.71\% & 0.5909 & \textbf{0.6379*} & 7.95\% & 0.6983 & \textbf{0.7312*} & 4.71\% \\
SetRank & 0.6509 & \textbf{0.6711*} & 3.10\% & 0.6384 & \textbf{0.6538*} & 2.41\% & 0.5947 & \textbf{0.6137*} & 3.19\% & 0.7006 & \textbf{0.7164*} & 2.26\% \\
MIR & 0.7178 & \textbf{0.7241*} & 0.88\% & 0.7011 & \textbf{0.7078*} & 0.96\% & 0.6747 & \textbf{0.6837*} & 1.33\% & 0.7549 & \textbf{0.7597*} & 0.64\% \\
\bottomrule
\end{tabular}
}}
\label{tab:reranking}
% \vspace{-3}
\footnotesize \flushleft\hspace{0cm} $*$ denotes statistically significant improvement over backbone reranking models (t-test with $p$-value $<$ 0.05).
\vspace{-5pt}
% \Jianghao{underline the second best value}
\end{table*}
\subsection{Effectiveness Comparison}
\subsubsection{Improvement over Backbone Models (RQ1)} On \textbf{CTR prediction} task, we implement our proposed KAR upon 9 representative CTR models, and the results are shown in Table~\ref{tab:backbone_model}. From the table, we can have the following observations: (i) Applying KAR significantly improves the performance of backbone CTR models. For example, when using FiBiNet as the backbone model on MovieLens-1M, KAR achieves a 1.49\% increase in AUC and a 2.27\% decrease in LogLoss, demonstrating the effectiveness of incorporating open-world knowledge from LLMs into RSs. (ii) As a model-agnostic framework, KAR can be applied to various types of baseline models, whether focusing on feature interaction or behavior modeling. With the equipment of KAR, the selected 9 representative CTR models on two datasets all achieve an AUC improvement of about 1-1.5\%, indicating the universality of the KAR. 
\textbf{(iii)} KAR shows more remarkable improvement in feature interaction models compared to user behavior models. This may be because the knowledge augmented vectors generated by KAR are utilized more effectively by the feature interaction layer than the user behavior modeling layer. The dedicated feature interaction design may better exploit the information contained in the knowledge vectors.

To investigate KAR's compatibility on other tasks, \eg, \textbf{reranking}, we incorporate KAR into the state-of-the-art reranking models. The results on the Amazon-Books dataset are presented in Table ~\ref{tab:reranking}, from which the following observations can be made: (i) KAR significantly enhances the performance of backbone reranking models. For example, when PRM is employed as the backbone, KAR achieves a remarkable increase of 5.71\% and 4.71\% in MAP@7 and NDCG@7. (ii) KAR demonstrates more pronounced improvements in methods that do not involve history modeling, such as DLCM, PRM, and SetRank. The user preference knowledge provided by KAR is particularly advantageous for these methods. However, in MIR, which adequately explores the relationship between history and candidates, the enhancement is slightly smaller.

\subsubsection{Improvement over Baselines (RQ2)} Next, we compare KAR with recent baselines using language models or sequence representation pretraining on CTR task. The results are presented in Table~\ref{tab:baselines}, from which we make the following observations: (i) KAR significantly outperforms models based on pretrained language models. For instance, with DIN as the backbone on Amazon-Books, KAR achieves a 0.91\% improvement in AUC and a 1.27\% improvement in LogLoss over the strongest baseline TALLRec. 
(ii) When integrating PLMs into RSs, the improvements brought by smaller PLMs are relatively modest. For instance, UniSRec, VQ-Rec, and P5, based on PLMs with parameters less than one billion, tend to exhibit poorer performance, even worse than the baseline DIN. (iii) Conversely, leveraging LLMs can lead to more substantial improvements, as seen in TALLRec and KAR. However, achieving effective enhancements with LLM embedding as initialization proves challenging, \eg, there is little difference between the results of LLM2DIN and DIN.

% (ii) Pretraining models such as UniSRec, VQ-Rec and P5, perform poorly, and even fail to surpass the basic DIN. This could be attributed to the fact that the those studies has little exploration of the feature interaction, while the CTR prediction heavily relies on feature interaction.

% ~\footnote{We did not include the results of P5, because when adapting its official code to CTR task, its metrics are notably worse than others even after careful parameter tuning. \xx{For instance, on MovieLens-1M, the AUC for P5 is only 0.713, significantly worse than other baselines. This may be attributed to the fact that CTR tasks heavily rely on feature interaction, while language models like P5 struggle to capture this effectively.}}
% (iii) P5 (to be completed)
\begin{table}[h]
\centering
% \vspace{-5pt}
\caption{Comparison between KAR and baselines. }
% \vspace{-5pt}
\scalebox{0.95}{
\setlength{\tabcolsep}{1.2mm}{
\begin{tabular}{cccc|cc}
\toprule
\multirow{2}{*}{\textbf{Model}} & \multirow{2}{*}{\textbf{\begin{tabular}[c]{@{}c@{}}Backbone\\ PLM\end{tabular}}}& \multicolumn{2}{c|}{\textbf{MovieLens-1M}} & \multicolumn{2}{c}{\textbf{Amazon-Books}} \\
\cmidrule{3-6}
 & & \textbf{AUC} & \textbf{LogLoss} & \textbf{AUC} & \textbf{LogLoss} \\
 \midrule
% UniSRec & BERT-110M & 0.7891 & 0.5496 & 0.8196 & 0.5063 \\
% VQ-Rec & BERT-110M & 0.7914 & 0.5456 & 0.8226 & 0.5025 \\
% P5& T5-223M & 0.7935 & 0.5436 & 0.8333 & 0.4908 \\
% LLM2DIN & ChatGLM-6B & 0.7977 & 0.5379 & 0.8307 & 0.4930  \\
% TALLRec & LLaMA2-7B & \underline{0.8007} & \underline{0.5372} & \underline{0.8342} & \underline{0.4862}  \\
% base(DIN) &N/A & 0.7975 & 0.5387 & 0.8304 & 0.4937 \\
% KAR(DIN) &gpt-3.5-turbo & \textbf{0.8066*} & \textbf{0.5304*} & \textbf{0.8418*} & \textbf{0.4801*} \\
UnisRec                         & BERT-110M                                                                         & 0.7702              & 0.5641              & 0.8196              & 0.5063              \\
VQ-Rec                          & BERT-110M                                                                         & 0.7707              & 0.5641              & 0.8226              & 0.5025              \\
P5                              & T5-223M                                                                           & 0.7790              & 0.5543              & 0.8333              & 0.4908              \\
LLM2DIN                         & ChatGLM-6B                                                                        & 0.7874              & 0.5473              & 0.8307              & 0.4930              \\
TALLRec                         & LLaMa2-7B                                                                         & \underline{0.7892}        & \underline{0.5451}        & \underline{0.8342}        & \underline{0.4862}        \\
base(DIN)                       & N/A                                                                               & 0.7863              & 0.5486              & 0.8304              & 0.4937              \\
KAR(DIN)                        & ChatGLM-6B                                                                        & \textbf{0.7961*}    & \textbf{0.5370*}    & \textbf{0.8418*}    & \textbf{0.4801*}   \\
\bottomrule
\end{tabular}
}}
\label{tab:baselines}
\footnotesize \flushleft\hspace{0cm} $*$ denotes statistically significant improvement over the second best baselines which is underlined (t-test with $p$-value $<$ 0.05).
\vspace{-5pt}
% \Jianghao{underline the second best value}
\end{table}

% \subsubsection{Generalization on Other Task (RQ3)} 

\subsubsection{Improvement over Other Knowledge (RQ3)} 
\begin{figure}[h]
    \centering
% \vspace{-7pt}
    \includegraphics[width=0.48\textwidth]{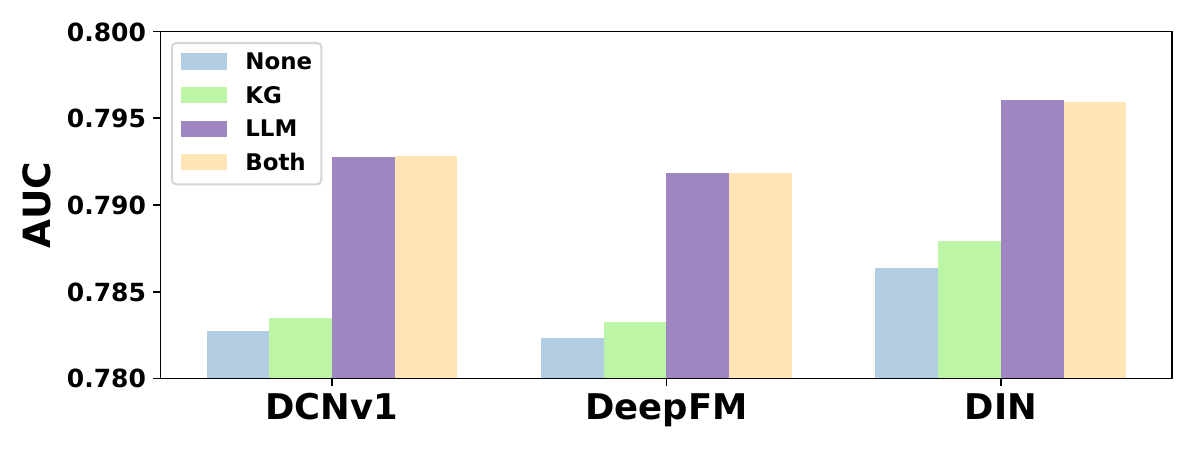}
% \vspace{-6pt}
    \caption{Comparison between knowledge from knowledge graph and LLM on MovieLens-1M dataset.}
\vspace{-5pt}
    \label{fig:kge}
\end{figure}

Finally, in Figure~\ref{fig:kge}, we compare knowledge from LLMs and other sources, such as knowledge graph (KG),  with DCNv1, DeepFM, and DIN as the backbone on MovieLens-1M dataset. We utilize a knowledge graph from LODrecsys~\cite{DOTD16}, which maps items of MovieLens-1M to DBPedia entities. Then, the entity embedding of each item is extracted following KTUP~\cite{cao2019unifying} and used as an additional feature for the backbone model. The legend "\textbf{None}" denotes the backbone model without knowledge enhancement, while "\textbf{KG}", "\textbf{LLM}", and "\textbf{Both}" represent the backbone model enhanced by knowledge from KG, LLM, and both sources, respectively.

From Figure~\ref{fig:kge}, we notice that knowledge from both KG and LLM can bring performance improvements; however, the enhancement of KG is much smaller compared to that of LLM. This might be attributed to the fact that KG only possesses manually annotated item-side knowledge while it lacks reasoning knowledge for user preferences. According to our analysis in the next subsection~\ref{reas_fact_konwledge}, reasoning knowledge for user preference usually contributes more gains compared to item factual knowledge. Furthermore, the simultaneous utilization of the two kinds of knowledge does not exhibit significant improvement over using LLM alone. This suggests that LLM may already encompass the typical knowledge within KG, making knowledge derived solely from LLM sufficient. 

\subsection{Deployment \& Online A/B Test (RQ4)}\label{sec:online}
To validate the effectiveness of KAR, we conducted two online experiments on Huawei's news and music platforms, respectively. On the news platform, the experimental group, where we deployed KAR, utilized Huawei's large language model PanGu\cite{zeng2021pangu} to generate user preference, followed by retrieving relevant news through vector similarity. The control group utilized the original recall model as our baseline. During the online A/B test, KAR exhibited a 7\% improvement on Recall metric compared with the baseline, resulting in significant business benefits. 
% We randomly divided 50\% of users into experimental group, and the remaining 50\% were assigned to the control group. The experimental group used knowledge vectors generated by KAR for recall, while the control group utilized the original recall model as our baseline. 

In the music scenario, 10\% of users were randomly selected into the experimental group and another 10\% were in the control group. Both groups used the same base model for generating recommendations. The difference lies in their inputs, where the input of the experimental group included the knowledge representation augmented by KAR.  To enhance the storage efficiency, we also employed PCA to reduce the dimension of representation to 64. In a 7-day online A/B test, KAR demonstrated a 1.7\% increase in song play count, a 1.64\% increase in the number of devices for song playback, and a 1.57\% increase in total duration. This indicates that KAR can be successfully implemented in industrial settings and improve recommendation experience for real-world users. 

% \ww{This part is too short, could add a flowchart for online deployment.}
% The experimental group received recommendation generated by a base model with vectors augmented by LLM, i.e., KAR, while the control group did not incorporate this enhancement.

% \vspace{-7pt}
\subsection{Ablation Study}
\subsubsection{Reasoning and Factual Knowledge (RQ5)}\label{reas_fact_konwledge}

\begin{figure}[h]
    \centering
% \vspace{-7pt}
    \includegraphics[width=0.48\textwidth]{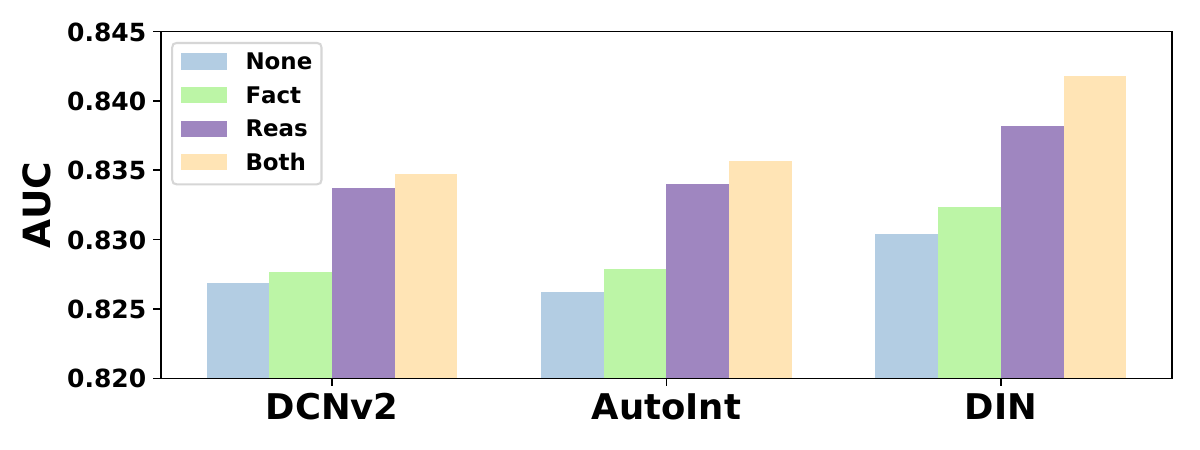}
% \vspace{-6pt}
    \caption{Ablation study about reasoning and factual knowledge on Amazon-Books dataset.}
\vspace{-5pt}
    \label{fig:knowledge_abaltion}
\end{figure}

To study the impact of knowledge generated by LLMs, we conduct an ablation study on reasoning and factual knowledge on the Amazon-Books dataset. We select DCNv2, AutoInt, and DIN as backbone models and compare their performance with different knowledge enhancements, as shown in Figure~\ref{fig:knowledge_abaltion}. The legend \textbf{"None"} represents the backbone model without any knowledge enhancement. \textbf{"Fact"} and \textbf{"Reas"} indicate the backbone models enhanced with factual knowledge on item and reasoning knowledge about user preference, respectively, while \textbf{"Both"} represents the joint use of both knowledge types.

From Figure~\ref{fig:knowledge_abaltion}, we observe that both reasoning knowledge and factual knowledge can improve the performance of backbone models, with reasoning knowledge exhibiting a larger improvement. This could be attributed to the fact that reasoning knowledge inferred by the LLMs captures in-depth user preferences, thus compensating for the backbone model's limitations in reasoning underlying motives and intentions. Additionally, the joint use of both reasoning and factual enhancements outperforms using either one alone, even achieving a synergistic effect where $1+1>2$. One possible explanation is that reasoning knowledge contains external information that is not explicitly present in the raw data. When used independently, this external knowledge could not be matched with candidate items. However, combining the externally generated factual knowledge on items from the LLMs aligned with reasoning knowledge allows RSs to gain a better understanding of items according to user preferences.

\begin{table}[h]
\centering
% \vspace{-7pt}
\caption{Impact of different knowledge encoders on MoiveLens-1M dataset. 
% LL is short for LogLoss.
% \Jianghao{The best results are given in bold, while the second-best values are underlined.}
}
% \vspace{-5pt}
\scalebox{1.0}{
\setlength{\tabcolsep}{1.2mm}{
\begin{tabular}{ccc|cc}
\toprule
\multirow{2}{*}{\textbf{Variants}}  & \multicolumn{2}{c}{\textbf{BERT}} & \multicolumn{2}{c}{\textbf{ChatGLM}} \\
 \cmidrule{2-5}
 & \textbf{AUC} & \textbf{LogLoss} & \textbf{AUC} & \textbf{LogLoss} \\
 \midrule
% base(DIN) & 0.7975 & 0.5387 & 0.7975 & 0.5387  \\
% KAR(LR) & 0.7746 & 0.5699 & 0.7903 & 0.5490  \\
% KAR(MLP) & 0.7882 & 0.5518 & 0.7986 & 0.5399 \\
% KAR-MLP & {\underline{0.8040}} & 0.5345 & 0.8042 & 0.5318  \\
% KAR-MoE & \textbf{0.8046} & \textbf{0.5322} & {\underline{0.8052}} & {\underline{0.5305}}  \\
% KAR & \textbf{0.8046} & {\underline{0.5324}} & \textbf{0.8066} & \textbf{0.5304} \\
base(DIN) & 0.7863 & 0.5486 & 0.7863 & 0.5486 \\
KAR(LR) & 0.7589 & 0.5811 & 0.7720 & 0.5674 \\
KAR(MLP) & 0.7754 & 0.5588 & 0.7816 & 0.5559 \\
KAR-MLP & 0.7934 & 0.5411 & 0.7939 & 0.5401 \\
KAR-MoE & \underline{0.7946} & \textbf{0.5391} & \underline{0.7950} & \underline{0.5394} \\
KAR & \textbf{0.7947} & \underline{0.5402} & \textbf{0.7961} & \textbf{0.5370} \\
\bottomrule
\end{tabular}
}}
% \Jianghao{(1) underline the second best value. (2) significant test is needed}
% \vspace{-5pt}
\label{tab:text_encoder_ml}
\end{table}

\begin{table}[h]
\centering
% \vspace{-7pt}
\caption{Impact of different knowledge encoders on Amazon-Books dataset. 
% LL is short for LogLoss.
% \Jianghao{The best results are given in bold, while the second-best values are underlined.}
}
% \vspace{-5pt}
\scalebox{1.0}{
\setlength{\tabcolsep}{1.2mm}{
\begin{tabular}{ccc|cc}
\toprule
\multirow{2}{*}{\textbf{Variants}} & \multicolumn{2}{c}{\textbf{BERT}} & \multicolumn{2}{c}{\textbf{ChatGLM}}  \\
 \cmidrule{2-5}
 & \textbf{AUC} & \textbf{LogLoss} & \textbf{AUC} & \textbf{LogLoss} \\
 \midrule
base(DIN) & 0.8304 & 0.4937 & 0.8304 & 0.4937 \\
KAR(LR)  & 0.7375 & 0.5861 & 0.7560 & 0.5718 \\
KAR(MLP)  & 0.7424 & 0.5834 & 0.7571 & 0.5763 \\
KAR-MLP & 0.8357 & 0.4880 & 0.8370 & 0.4843 \\
KAR-MoE & {\underline{0.8371}} & \textbf{0.4841} & {\underline{0.8388}} & {\underline{0.4823}} \\
KAR & \textbf{0.8374} & {\underline{0.4843}} & \textbf{0.8418} & \textbf{0.4801}\\
\bottomrule
\end{tabular}
}}
% \Jianghao{(1) underline the second best value. (2) significant test is needed}
% \vspace{-5pt}
\label{tab:text_encoder_amz}
\end{table}

\subsubsection{Knowledge Encoders and Semantic Transformation (RQ6)}\label{sec:knowledge_encoder}
We employ two different language models, BERT~\cite{bert} and ChatGLM~\cite{du2022glm}, to investigate the impact of different knowledge encoders on model performance. Additionally, we design several variants to demonstrate how the representations generated by knowledge encoders are utilized. \textbf{KAR(LR)} applies average pooling to token representations from the knowledge encoder and directly feeds the result into a linear layer to obtain prediction scores, without utilizing a backbone CTR model. \textbf{KAR(MLP)} replaces the linear layer of KAR(LR) with an MLP. \textbf{KAR-MLP} and \textbf{KAR-MoE} replace the hybrid-expert adaptor with an MLP and a Mixture-of-Experts (MoE), respectively. The original KAR and the two variants, KAR-MLP and KAR-MoE, all adopt DIN as the backbone model. Table~\ref{tab:text_encoder_ml} and~\ref{tab:text_encoder_amz} presents their performance, from which we draw the following conclusions. 

Firstly, we can observe that, overall, variants with ChatGLM as the knowledge encoder outperform those with BERT. The performance of KAR(LR) and KAR(MLP) can be considered as a measure of the quality of the encoded representations, since they directly adopt the representations for prediction. Considering KAR(LR) and KAR(MLP), 
the superior performance of ChatGLM over BERT indicates that ChatGLM performs better in preserving the information within knowledge from LLMs, which may be attributed to the larger size and better text comprehension of ChatGLM (6 billion) compared to BERT (110 million). With ChatGLM on MovieLens-1M, KAR(MLP) is even close to some base CTR model like DeepFM in Table~\ref{tab:backbone_model}, validating the effectiveness of our generated open-world knowledge.

% Since KAR(LR) and KAR(MLP) only utilize the knowledge representations from knowledge encoder, their results can be considered as a measure of the information contained in the representations. On the two variants, 

% Secondly, only leveraging the open-world knowledge generated by LLMs for recommendation can lead to some improvements in certain cases, yet integrating it with recommendation domain knowledge yields more promising results. Among the methods directly utilizing encoded representations, only KAR(MLP) with ChatGLM gains a modest improvement (0.14\% in AUC) over the base(DIN) on MovieLens-1M. However, combining the generated open-world knowledge with the domain knowledge in the classical RSs, as in KAR, brings significant enhancements (1.15\% and 1.38\% AUC improvement) on two datasets. We attribute this improvement to KAR successfully bridging the open-world knowledge and the recommendation domain knowledge.

Secondly, the performance of KAR benefits from complex semantic transformation structures, but the knowledge encoder also limits it. The results on ChatGLM show that a simple MLP is less effective than MoE and our designed hybrid-expert adaptor outperforms the MoE, indicating that the transformation from semantic space to recommendation space entails a complex network structure. However, with BERT as knowledge encoder, KAR-MoE and KAR-MLP exhibit similar performance, suggesting that information from BERT is limited and using an MoE is sufficient in this case.
\vspace{-5pt}
\subsection{Efficiency Study (RQ7)}\label{sec:efficiency}
To quantify the actual time complexity of KAR, we compare the inference time of KAR based on DIN with a widely-used LLM API, strongest baseline TALLRec, and base DIN model in Table~\ref{tab:inference}. For \textbf{LLM API}, we follow the zero-shot user rating prediction in~\cite{kang2023llms} that provides the user viewing history and ratings as prompt and invokes the API to predict user ratings on candidate items. Since this approach does not allow setting a batch size, the table presents the average response time per sample. For KAR, we evaluate the two acceleration strategies as introduced in Section~\ref{sec:speed_up}: $\textbf{KAR}_{w/\,apt}$, where the adaptor participates in the inference stage, and $\textbf{KAR}_{w/o\,apt}$, where the adaptor is detached from inference. The experiments of KAR and base model are all conducted on a Tesla V100 with 32G memory, with a batch size of 256. With the same Tesla V100, we also test \textbf{TALLRec} and only showcase the average inference time per sample, since 32G memory cannot handle LLM with batch size of 256. Table~\ref{tab:inference} presents the average inference time, from which we draw following conclusions.

Firstly, adopting LLM for direct inference is not feasible for RSs due to its large computational latency. The response latency of LLM API is 4-6 seconds, which does not meet the real-time requirement of RSs typically demanding a response latency of within 100ms. Even if we finetune a relatively small LLM, such as TALLRec based on LLaMa2-7B, its inference latency is still close to 1s, which is unbearable for industrial scenarios. Secondly, both acceleration methods of KAR achieve an inference time within 100ms, satisfying the low latency requirement. Importantly, if we employ the approach of prestoring reasoning and factual augmented vectors, \ie, $\text{KAR}_{w/o\,apt}$, the actual inference time is nearly the same as that of the backbone model. This demonstrates the effectiveness of our proposed KAR and acceleration strategies.

\begin{table}[h]
\centering
% \vspace{-5pt}
\caption{The comparison of inference time (s).}
% \vspace{-5pt}
\scalebox{1.0}{
\setlength{\tabcolsep}{1.2mm}{
\begin{tabular}{ccc}
\toprule
\textbf{Model}      & \textbf{MovieLens-1M} & \textbf{Amazon-Books} \\
\midrule
LLM API   & $5.54$ & $4.11$ \\
TALLRec   &  $7.63\times10^{-1}$ & $7.97\times10^{-1}$ \\
% P5   &  &  \\
$\text{KAR}_{w/\,\,apt}$  & $8.08\times10^{-2}$ & $9.39\times10^{-2}$ \\
$\text{KAR}_{w/o\,\,apt}$ & $6.64\times10^{-3}$ & $1.11\times10^{-2}$\\
base DIN     & $6.42\times10^{-3}$ & $1.09\times10^{-2}$\\
\bottomrule
\end{tabular}
}}
\vspace{-5pt}
\label{tab:inference}
\end{table}

\section{Broader Impact}
When incorporating LLMs into RSs, it is imperative to give serious consideration to privacy and security issues. While LLMs offer remarkable capabilities, they also come with potential risks of compromising user privacy and generating harmful content~\cite{pan2020privacy,Brown2022privacy,weidinger2021ethical}. We also consider these two issues while designing our proposed framework, KAR.

Regarding privacy, KAR leverages LLMs' reasoning ability and factual knowledge without finetuning LLMs, ensuring that LLMs do not retain or remember user-specific data. Besides, while experiments may involve the use of external APIs on public datasets, real-world implementations usually rely on in-house models, \eg, we utilize Huawei's own LLM PanGu\cite{zeng2021pangu} for online A/B test as described in section~\ref{sec:online}, preventing the leakage of user privacy information through external APIs.

Furthermore, compared to methods that directly display LLM-generated content to users~\cite{chatrec,liu2023chatgpt,dai2023uncovering}, KAR takes a  more proactive stance to mitigate concerns about harmful content and hallucination knowledge generated by LLMs. KAR first converts the textual content from LLMs to robust representations and then incorporates those representations into traditional RSs. This integration avoids displaying
harmful or misleading content generated by LLMs while enabling us to utilize filtering mechanisms commonly used in traditional RSs to screen out potential harmful item recommendations.

Through those measures, KAR strives to deliver robust recommendation performance while safeguarding user data privacy and the security of recommendation content. 
% This approach is aimed at maintaining user trust ans ensuring a high-quality personalized recommendation experience.

\section{Conclusion}

Our work presents KAR, a framework for effectively incorporating the open-world knowledge into recommender systems by exploiting large language models. KAR identifies two types of critical knowledge from LLMs, the reasoning knowledge on user preferences and the factual knowledge on items, which can be proactively acquired by our designed factorization prompting. A hybrid-expert adaptor is devised to transform the obtained knowledge for compatibility with recommendation tasks. The obtained augmented vectors can then be used to enhance the performance of any recommendation model. Additionally, efficient inference is achieved through preprocessing and prestoring the LLM knowledge. KAR shows superior performance compared to the state-of-the-art methods and is compatible with various recommendation algorithms. 
% \Jianghao{maybe we should mention some future works}

% In future work, we plan to explore additional tasks, datasets, and diverse strategies for utilizing augmented features. 

% \balance
\bibliographystyle{ACM-Reference-Format}
\bibliography{reference}
% \clearpage
% \input{appendix}
\end{document}